\documentclass[aps,pre,twocolumn,showpacs,groupedaddress]{revtex4}

\usepackage{graphicx}
\usepackage{dcolumn}
\usepackage{bm}
\usepackage{amssymb}

\begin{document}


\title{Dynamic Modes of Microcapsules in Steady Shear Flow: Effects of Bending and Shear Elasticities}

\author{Hiroshi Noguchi}
\email[]{noguchi@issp.u-tokyo.ac.jp}
\affiliation{
Institute for Solid State Physics, University of Tokyo,
 Kashiwa, Chiba 277-8581, Japan}

\date{\today}

\begin{abstract}
The dynamics of microcapsules in steady shear flow was studied using a 
theoretical approach based on three variables: The Taylor deformation parameter $\alpha_{\rm D}$,  
the inclination angle $\theta$, and the phase angle $\phi$ of the membrane rotation.
It is found that the dynamic phase diagram shows a remarkable change with an increase in
the ratio of the membrane shear and bending elasticities.
A fluid vesicle (no shear elasticity) exhibits three dynamic modes: (i) Tank-treading (TT)
at low viscosity $\eta_{\rm {in}}$ of internal fluid  
($\alpha_{\rm D}$ and $\theta$ relaxes to constant values),
(ii) Tumbling (TB) at high $\eta_{\rm {in}}$ ($\theta$ rotates), and
(iii) Swinging (SW) at middle $\eta_{\rm {in}}$ and high shear rate $\dot\gamma$ ($\theta$ oscillates).
All of three modes are accompanied by a membrane ($\phi$) rotation.
For microcapsules with low shear elasticity,
the TB phase with no $\phi$ rotation and the coexistence phase of SW and TB motions
 are induced by the energy barrier of $\phi$ rotation.
Synchronization of $\phi$ rotation with TB rotation or SW oscillation occurs 
with integer ratios of rotational frequencies.
At high shear elasticity, where a saddle point in the energy potential disappears,
intermediate phases vanish, and either $\phi$ or $\theta$ rotation occurs. 
This phase behavior agrees with recent simulation results of microcapsules with low bending elasticity.
\end{abstract}
\pacs{87.16.D-, 05.45.-a, 82.40.Bj}

\maketitle

\section{Introduction}

Soft deformable objects such as liquid droplets~\cite{rall84,ston94}, 
lipid vesicles~\cite{haas97,abka05,made06,kant05,kant06,desc09,desc09a,kant07,krau96,seif99,misb06,vlah07,dank07,lebe07,lebe08,turi08,nogu04,nogu05,nogu07b,nogu09,nogu10,nogu10a}, 
red blood cells (RBCs)~\cite{skal90,fung04,fisc78,abka07,abka08a,kell82,tran84,naka90,wata06,pozr03,pozr05,skot07,dupi07,macm09,nogu05b,mcwh09,nogu09b}, 
and synthetic capsules~\cite{chan93,walt01,kess08,sui08,bagc09,kess09,lac08,lefe08,dods08,tsub10}
exhibit rich behaviors in flows.
In recent years, they have received growing attention experimentally, theoretically, and numerically.
These objects are filled with liquids and have a constant internal volume $V$.
The lipid vesicles, RBCs, and synthetic capsules are wrapped by lipid or polymer membranes
and resist  breaking up under much larger shear than liquid droplets.
The variation of material properties of these objects
can lead to different dynamic modes with morphological changes.

Since microcapsules can carry substances and control their release,
they are widely used in various fields including food, printing inks, pharmaceutics (drug carrier for drug delivery system),
agrochemicals, and cosmetics. 
Various kinds of microcapsules are synthesized with different sizes and 
membrane thicknesses $h$ \cite{chan93,walt01,elsn04,saka08,yow09}.
The deformability of microcapsules can be varied widely by controlling the membrane elastic properties
and capsule shapes.
When the membrane is made of linear-elastic material,
two-dimensional Young's modulus $Y_{\rm {mb}}\equiv 4\mu K_{\rm A}/(\mu+K_{\rm A})=Yh$ and
the bending modulus $\kappa = Yh^3/12(1-\nu^2)$,
where $\mu$ and $K_{\rm A}$ are two-dimensional shear and area compression moduli, respectively,
with three-dimensional Young's modulus $Y$ and Poisson ratio $\nu$ \cite{land86}.
Thus, thicker membranes become more difficult to bend than to stretch ($\kappa/Y_{\rm {mb}} \propto h^2$).

RBCs have a composite membrane made of lipid membrane and attached spectrin network.
The lipid membrane is in a fluid phase and is  almost incompressible.
The spectrin network adds  shear elasticity to the membrane.
The bending and shear elasticities of RBCs are estimated by 
micropipette  aspiration \cite{fung04,moha94} and 
optical tweezers experiments \cite{leno01,mill04,dao06}: $\kappa\simeq 2 \times 10^{-19}$J
and $\mu \simeq 6 \times 10^{-6}$N/m.

The relative importance of bending and shear elasticities is determined by the 
dimensionless F\"oppl-von K\'arm\'an number
$\gamma=Y_{\rm {mb}} R_0^2/\kappa$, where 
$R_0$ is the mean radius of capsules or cells \cite{lidm03}.  
The buckling transition of capsules from spherical to icosahedral shape occurs at $\gamma_{\rm {b}}\simeq 150$ \cite{lidm03}.
Capsules with a linear elastic membrane have large values of $\gamma \gtrsim 100$
because $\gamma= 12(1-\nu^2) (R_0/h)^2$.
Capsules with a composite membrane
can also have small $\gamma$. In this paper, we investigate microcapsules with wide ranges of $\gamma$.

\begin{figure}
\includegraphics{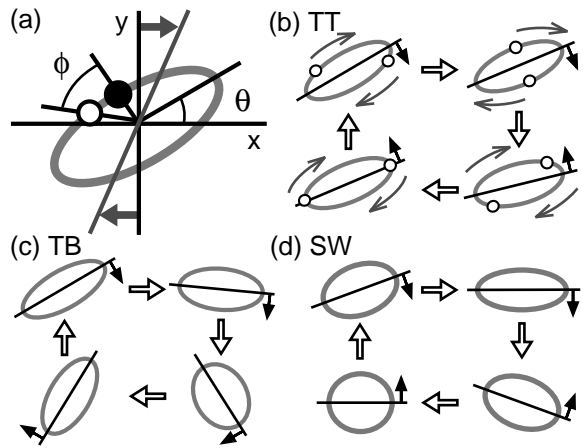}
\caption{ \label{fig:cart}
Schematic representation of (a) a capsule and its dynamic modes:
(b) Tank-treading (TT), (c) Tumbling (TB), and (d) Swinging (SW) modes.
(a) The inclination angle $\theta$ and phase angle $\phi$ are depicted.
The closed and open circles are the original and current positions of a membrane segment, respectively.
(b) For an elastic capsule, a capsule shape and $\theta$ oscillate,
while they are kept at fixed values for a fluid vesicle.
}
\end{figure}

Recently, the dynamics of lipid vesicles in steady shear flow was intensively 
investigated~\cite{haas97,abka05,made06,kant05,kant06,desc09,desc09a,kant07,krau96,seif99,misb06,dank07,lebe07,lebe08,turi08,vlah07,nogu04,nogu05,nogu07b,nogu09}.
A lipid vesicle can be considered as a microcapsule in the small limit of the shear modulus $\mu\to 0$.
An isolated fluid vesicle exhibits three types of dynamic modes in a steady shear flow, 
with flow velocity ${\bf v}=\dot\gamma y {\bf e}_x$, 
where ${\bf e}_x$ is a unit vector in the flow direction.
When the viscosity of internal fluid $\eta_{\rm {in}}$ and membrane viscosity $\eta_{\rm {mb}}$
are low, the membrane exhibit a tank-treading (TT) rotation with a stationary 
shape and a constant inclination angle $\theta>0$ [see Fig. \ref{fig:cart}(b)].
At high $\eta_{\rm {in}}$ or $\eta_{\rm {mb}}$,
the vesicle exhibit a tumbling (TB) motion, where $\theta$ rotates [see Fig. \ref{fig:cart}(c)].
Around the TT-TB transition viscosity with high shear rate $\dot\gamma$,
a swinging (SW) motion appears \cite{kant06,desc09,desc09a,misb06,dank07,lebe07,lebe08,nogu07b},
where $\theta$ and the vesicle shape oscillate [see Fig. \ref{fig:cart}(d)].
This motion is also called trembling \cite{kant06,desc09,desc09a,lebe07,lebe08} or vacillating-breathing \cite{misb06,dank07}.
These three types of motion can be understood by
the perturbation theories for quasi-spherical vesicles
\cite{dank07,lebe07,lebe08} or a generalized 
Keller-Skalak (KS) theory for deformable ellipsoidal vesicles \cite{nogu07b}. 

Experimentally, synthetic capsules and RBCs 
show the oscillation (swinging) of their lengths and $\theta$ \cite{chan93,walt01,abka07}
during TT motion and RBCs
also transit from TB to TT with increasing $\dot\gamma$  \cite{gold72,abka07}.
Recently, this dynamics was explained by the KS theory with the 
addition of an energy barrier for the TT rotation caused by the 
membrane shear elasticity~\cite{skot07,abka07}.
More recently, we phenomenologically extended this theory~\cite{skot07} to include the shape deformation of RBCs~\cite{nogu09b}.
Fix-shaped and deformable RBCs give very similar phase behavior. 
At low $\dot\gamma$, RBCs show TB ($\theta$) rotation without TT ($\phi$) membrane rotation.
In the middle ranges of shear rate $\dot\gamma$,
synchronized phases of the $\theta$ and $\phi$ rotations with integer ratios of the rotation frequencies
as well as intermittent rotations~\cite{skot07} appear~\cite{nogu09b}.
The TT-TB transition was also obtained by recent simulations of microcapsules~\cite{kess08,sui08,bagc09}.
However, the above intermediate phase (synchronized or intermittent rotations) was not observed in their simulations.
Kessler {\it et al.} contested that intermittent rotation is an artifact of the theoretical model~\cite{kess08,kess09}.
As we describe below, the disappearance of the intermediate phase is found to be caused 
by low bending elasticity in their simulations.
In this paper, we systematically investigate the dynamics of microcapsules in steady shear flow,
and show that their phase behaviors are qualitatively changed
from the bending-elasticity dominant condition  to the shear-elasticity dominant condition.

Note that some confusion is caused by the different definitions of a tumbling state.
For fluid vesicles, tumbling typically means  the rotation of inclination angle 
 accompanied by the rotation of the phase angle $\phi$.
On the other hand, for RBCs and capsules, it typically means the $\theta$ rotation
 accompanied by the $\phi$ oscillation (not rotation).
Thus, in the definition for fluid vesicles,
synchronized or intermittent rotations of $\theta$ and $\phi$ are also called ``tumbling''.
In this paper, we call them ``TB with no TT'' and ``TB with TT'' to distinguish their $\phi$ modes.

The dynamics of microcapsules is described by several dimensionless quantities.
Here, dimensionless quantities are denoted by a superscript $*$.
The relative ratio of the volume $V$ and surface area $S$ is characterized by
the reduced volume $V^*= 3V/(4\pi S)^{3/2} = (R_{\rm V}/R_{\rm S})^3= (1+\Delta_{\rm S}/4\pi)^{-3/2}$,
where $R_{\rm V}= (3V/4\pi)^{1/3}$, $R_{\rm S}=(S/4\pi)^{1/2}$, 
and the excess area $\Delta_{\rm S}=S/R_{\rm V}^2 -4\pi$.
The relative viscosity of the inside fluid and membrane are 
$\eta_{\rm {in}}^*=\eta_{\rm {in}}/\eta_0$
and $\eta_{\rm {mb}}^*=\eta_{\rm {mb}}/\eta_0 R_{\rm S}$, where 
$\eta_0$ is the viscosity of the outside fluid. 
The bending rigidity $\kappa$ is used as an energy unit, and
the shape relaxation time of the fluid vesicles with bending rigidity $\kappa$ is given
by $\tau=\eta_0 R_{\rm S}^3/\kappa$ (for $\eta_{\rm {in}}^*=1$). This
time is used to define a reduced shear rate $\dot\gamma^*=\dot\gamma \tau$. 
Since the Reynolds number is low, Re$<1$ in typical experimental conditions, 
the effects of the inertia are neglected.

The theoretical models of fluid vesicles and microcapsules are explained in Sec. \ref{sec:theory}.
The results are described in Sec. \ref{sec:results}.
First, the dynamics of fluid vesicles ($\mu=0$) is briefly explained in Sec. \ref{sec:rfluid},
and then dynamics with low and high shear elasticity $\mu$ is described in Secs. \ref{sec:rel1}--\ref{sec:min}.
The mechanism of the dynamic modes and experimental conditions are discussed in Sec. \ref{sec:dis}.
The summary is given in Sec. \ref{sec:sum}.

\section{Generalized Keller-Skalak Theory} \label{sec:theory}

\subsection{Fluid vesicle} \label{sec:thfluid}

Keller and Skalak \cite{kell82}
 analytically derived the equation of motion of the inclination angle $\theta$
for vesicles or elastic capsules
based on Jeffery's theory \cite{jeff22}.
The KS theory assumes (i) a fixed ellipsoidal shape of vesicles
and (ii) a simple velocity field on the membrane.
It gives very good predictions for vesicle dynamics at low shear rates.
The KS theory is phenomenologically extended to include the shape deformation of fluid vesicles 
 on the basis of the perturbation theory \cite{seif99,misb06,lebe07,lebe08} of quasi-spherical vesicles in Ref. {\cite{nogu07b}}.
The shape deformation is represented by the Taylor deformation parameter $\alpha_{\rm D}=(L_1-L_2)/(L_1+L_2)$, 
where $L_1, L_2$ are the principal lengths on the vorticity ($xy$) plane, and $L_3$ is  in the vorticity ($z$) direction.
The differential equations of $\alpha_{\rm D}$ and $\theta$ are given by 
\begin{eqnarray}
\label{eq:ald}
\frac{\ \ \ d \alpha_{\rm D}}{\dot\gamma dt} &=&  \left\{1-\left(\frac{\alpha_{\rm D}}
                   {\alpha_{\rm D}^{\rm {max}}}  \right)^2\right\}
    \left\{ -\frac{A_0}{\dot\gamma^*} 
       \frac{\partial F^*}{\partial \alpha_{\rm D}} 
                +  A_1\sin(2\theta)\right\}, \\
\label{eq:qks}
\frac{\ \ d\theta}{\dot\gamma dt} &=&  \frac{1}{2}\big\{-1+f_0 f_1 \cos(2\theta)\big\} - \frac{f_0d\phi}{\dot\gamma dt}, \\
\label{eq:phiks}
\frac{\ \ d \phi}{\dot\gamma dt} &=& -\frac{ \cos(2\theta) }
      {2f_1\{1+f_2(\eta_{\rm {in}}^* -1) 
                  + f_2f_3 \eta_{\rm {mb}}^*\}},
\end{eqnarray}
where $A_0= 45/8\pi(32+23\eta_{\rm {in}}^*+16\eta_{\rm {mb}}^*)V^*$ and 
$A_1= 30/(32+23\eta_{\rm {in}}^*+16\eta_{\rm {mb}}^*)$ \cite{nogu07b,nogu10}.
The first and second terms in the last parentheses of Eq.~(\ref{eq:ald}) represent
the forces of the membrane bending elasticity and the external shear stress, respectively.
The free energy is given by $F = F_{\rm {cv}} \equiv (\kappa/2) \int (C_1+C_2)^2 dS$,
where $C_1$ and $C_2$ are the principal curvatures at each point of the membrane. 

In Eq.~(\ref{eq:qks}),
the factors $f_0$, $f_1$, $f_2$, and $f_3$ are function of the vesicle shape ($\alpha_{\rm D}$, $V^*$) (see Appendix A).
Because the factors and $F$ are independent of the phase angle $\phi$
for fluid vesicles, Eq.~(\ref{eq:qks}) can be simply written as
\begin{equation} \label{eq:qks2}
\frac{\ \ d\theta}{\dot\gamma dt} = \frac{1}{2}\{-1+B\cos(2\theta)\}.
\end{equation}
The factor $B$ is a function 
of $\alpha_{\rm D}$, $V^*$, $\eta_{\rm {in}}^*$, and $\eta_{\rm {mb}}^*$.

At $\eta_{\rm {in}}^*=1$, the lipid vesicle shows only the TT motion at any $V^*$ and $\dot\gamma^*$.
The TT steady angle $\theta_0$ predicted by KS theory [Eq. (\ref{eq:qks})]
gives very good agreement with simulations and experiments \cite{krau96,nogu05,dupi07,abka05,kant05,nogu10}.
On the other hand, the perturbation theories 
give TB motion even at $\eta_{\rm {in}}^*=1$ for $V^* \lesssim 0.8$,
while they give good predictions for $V^* \gtrsim 0.9$ \cite{nogu10}.
Thus, the perturbation theories should not be applied 
to the dynamics of vesicles and capsules for $V^* \lesssim 0.8$.

\begin{figure}
\includegraphics{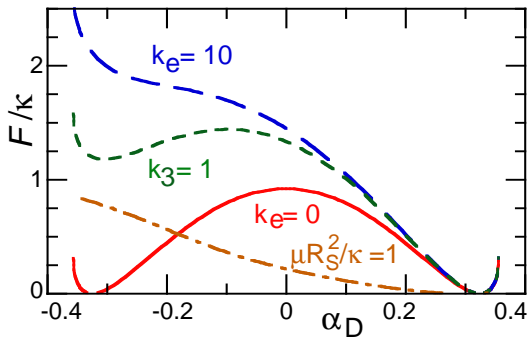}
\caption{ \label{fig:pot}
Free-energy potential function of the Taylor deformation parameter $\alpha_{\rm D}$ at $\phi=0$ and $\phi=\pi/2$ [$F(-\alpha_{\rm D},0)=F(\alpha_{\rm D}, \pi/2)$].
Solid and long-dashed lines represent $k_{\rm e}=0$ and $k_{\rm e}=10$ with $k_3=0$, respectively.
Short-dashed line represents $k_{\rm e}=10$ with $k_3=1$.
Dashed-doted line represents the potential of the shear elasticity estimated by the perturbation 
of quasi-spherical capsules with $\mu R_{\rm S}^2/\kappa=1$.
}
\end{figure}

\subsection{Microcapsule} \label{sec:thcap}

Synthetic capsules and RBCs have membrane shear elasticity $\mu$,
which modifies their free-energy landscape: $F=F_{\rm {cv}}+F_{\mu}$.
The energy of the shear elasticity $F_{\mu}$ depends on the capsule shape and phase angle $\phi$,
while $F_{\rm {cv}}$ only depends on the capsule shape.
The angle $\phi=0$ is the original position in the absence of flow.
During $\phi$ (TT) rotation, the membrane is locally elongated and
it generates a free-energy barrier at $\phi=\pi/2$.
Recently, Skotheim and Secomb extended the KS theory to take into account this free-energy barrier 
for capsules with fixed ellipsoidal shapes \cite{skot07}.
Equation (\ref{eq:phiks}) is modified as
\begin{equation} \label{eq:phiks2}
\frac{\ \ d \phi}{\dot\gamma dt} = -\frac{(c_0/\dot\gamma^*) \partial F^*/\partial \phi + \cos(2\theta) }
      {2f_1\{1+f_2(\eta_{\rm {in}}^* -1) 
                  + f_2f_3 \eta_{\rm {mb}}^*\}},
\end{equation}
where $c_0=3f_2/8\pi f_1 V^*$ \cite{skot07}.

In this paper, we employ Eqs. (\ref{eq:ald}), (\ref{eq:qks}), and (\ref{eq:phiks2})
to investigate the dynamics of the elastic capsule with prolate shape at $V^*=0.9$ and $\eta_{\rm {mb}}^*=0$.
These equations are numerically integrated using the fourth-order Runge-Kutta method.
The curvature energy $F_{\rm {cv}}(\alpha_{\rm D})$ is numerically calculated for the ellipsoidal vesicle.
The prolate ($L_1>L_2=L_3$) and oblate ($L_1=L_2>L_3$ and $\alpha_{\rm D}=0$) shapes are energy 
minimum and maximum, respectively (see Fig. \ref{fig:pot}).
Here, a fit function is employed to avoid the artifacts of nonsmooth functions:
$\partial F_{\rm {cv}}^*/\partial \alpha_{\rm D}= 30 \alpha_{\rm D} -820 \alpha_{\rm D}^5 -1.7/\sqrt{\alpha_{\rm D}^{\rm {max}}-\alpha_{\rm D}} +1.7/\sqrt{\alpha_{\rm D}^{\rm {max}}}$ with
$\alpha_{\rm D}^{\rm {max}}=0.356$.
Here, the relative energy to the free-energy minimum (prolate shape) is considered.
The force $-\partial F_{\rm {cv}}/\partial \alpha_{\rm D}$ diverges in the limit of maximum
extension, $\alpha_{\rm D} \to \alpha_{\rm D}^{\rm {max}}(V^*)$.
For shear elasticity, the leading-order energy function is employed in Secs. \ref{sec:rel1} and \ref{sec:rel5}:
 $F_{\mu}(\alpha_{\rm D},\phi)^*= (k_{\rm e}/2)\{(\alpha_{\rm D}- \alpha_0)^2\cos^2(\phi)+(\alpha_{\rm D}+ \alpha_0)^2\sin^2(\phi)\}$,
where $\alpha_0=0.325$ is $\alpha_{\rm D}$ of the prolate shape.
Because compression and $\pi/2$ rotation can generate the same capsule state,
the free energy should keep the relation $F(-\alpha_{\rm D}, \phi)=F(\alpha_{\rm D}, \phi+\pi/2)$.
The shear elastic energy is estimated by the perturbation of quasi-spherical capsules; see Appendix B.
The harmonic potential $F_{\mu}^*(\alpha_{\rm D},0)$ with $k_{\rm e} \simeq 4\mu R_{\rm S}^2/\kappa$
gives good agreement with the perturbation result; see Fig. \ref{fig:pot}.
In Sec. \ref{sec:min}, in order to check dependence on the potential shape, a higher-order term $F_3$ is added to 
the potential: $F_3^*=(k_{\rm e}k_3/3)\{(\alpha_{\rm D}- \alpha_0)^3\cos^2(\phi)-(\alpha_{\rm D}+ \alpha_0)^3\sin^2(\phi)\}$.

\section{Results} \label{sec:results}

\subsection{Fluid vesicle} \label{sec:rfluid}

\begin{figure}
\includegraphics{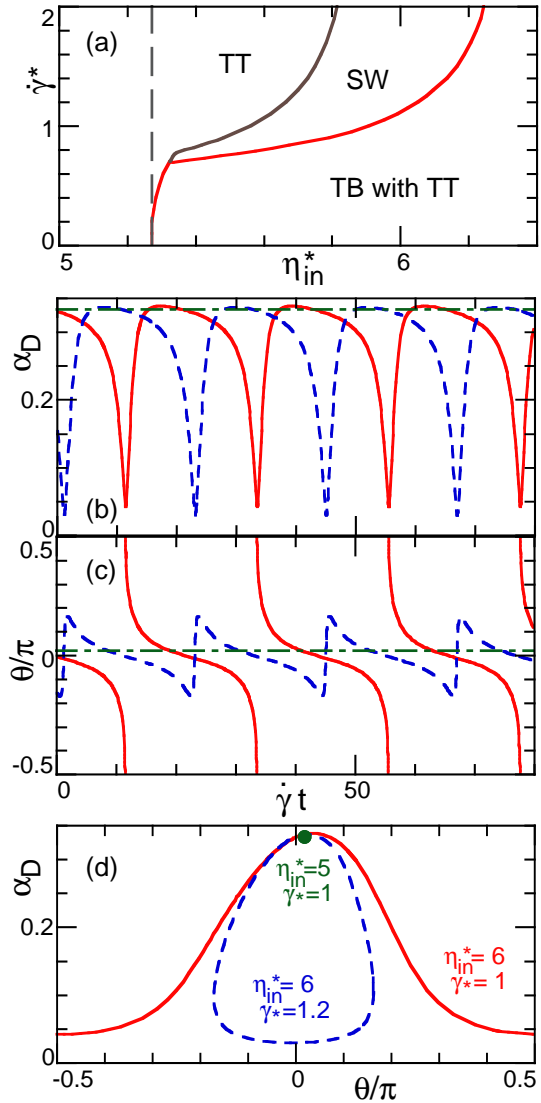}
\caption{ \label{fig:phase_e0}
(Color online)
Dynamics of fluid vesicles ($k_{\rm e}=0$).
(a) Phase diagram ($\eta_{\rm {in}}^*,\dot\gamma^*$).
Solid (red and brown) lines represent phase boundaries between
TB, TT, and SW. 
For the right region of the dashed line, the vesicle passes through or remains at $\theta<0$.
(b--d) Time development of $\alpha_{\rm D}$ and $\theta$.
Solid and dashed lines represent the TB and SW motions at ($\eta_{\rm {in}}^*,\dot\gamma^*)=(6,1)$ and $(6,1.2)$, respectively. 
Dashed-dotted lines and closed circle represent the TT motion at ($\eta_{\rm {in}}^*,\dot\gamma^*)=(5,1)$.
}
\end{figure}

First, we briefly describe the dynamics of lipid vesicles ($k_{\rm e}=0$) in steady shear flow.
Figure \ref{fig:phase_e0}(a) shows its phase diagram.
At low viscosity ratio $\eta_{\rm {in}}^*$, 
the vesicle shows a TT motion, where $\alpha_{\rm D}$ and $\theta$ remain constant.
The factor $B$ in Eq.(\ref{eq:qks2}) is more than unity, and
 a stable fixed point $\theta=\theta_0=0.5\arccos(1/B)$ 
and unstable fixed point $\theta=-\theta_0$ exist.
As $\eta_{\rm {in}}^*$ increases, 
$B$ and $\theta_0$ decrease.
At the critical viscosity $\eta_{\rm {in}}^*=\eta_{\rm {c}}^*=5.27$,
$B$ becomes unity and 
 the two fixed points $\theta=\pm\theta_0$ merge at $\theta=0$ (saddle-node bifurcation).
At $\eta_{\rm {in}}^*>\eta_{\rm {c}}^*$ with low shear rate $\dot\gamma^* \lesssim 1$,
no fixed point exists so
 $\theta$ rotates (TB with TT); see Figs. \ref{fig:phase_e0}(b--d).
This TT-TB transition is quantitatively predicted by the original KS theory for fix-shaped vesicles.

On the other hand, at higher shear $\dot\gamma^* \gtrsim 1$,
the shape deformation of the vesicles is not negligible.
A SW phase appears between the TT and TB phases.
In the SW phase, $\theta$ and $\alpha_{\rm D}$ oscillate; see dashed lines in Figs. \ref{fig:phase_e0}(b--d).
These oscillations are generated by the shape deformation  as follows \cite{nogu07b}:
In Eq.~(\ref{eq:ald}), the shear force depends on $\theta$ as $\sin(2\theta)$,
so the shear increases $\alpha_{\rm D}$ (elongation) for $0<\theta<\pi/2$,
but decreases  $\alpha_{\rm D}$ (shrinkage) for $-\pi/2<\theta<0$.
A prolate vesicle starts $\theta$ rotation with $B<1$ as in the TB phase
and then it shrinks to a more spherical shape at $\theta<0$,
which has greater $B$.
When $B$ becomes greater than $1/\cos(2\theta)$,
the right-hand side of Eq.(\ref{eq:qks2}) changes its sign and
the angle $\theta$ increases.
At $\theta>0$, the vesicle elongates back to the prolate shape.

The TB-SW transition occurs when the trajectory crosses $\alpha_{\rm D}=0$.
Between $\eta_{\rm {in}}^*=\eta_{\rm {c}}^*$ (dashed line in Fig. \ref{fig:phase_e0}(a)) 
and the SW phase, the TT phase has negative steady angle $\theta_0<0$.
For fluid vesicles,  the phase angle $\phi$ always shows rotation except for the limit $\eta_{\rm {in}}^* \to\infty$.
At $\eta_{\rm {in}}^* \to\infty$,
Eq.~(\ref{eq:qks2}) becomes  Jeffery's equation \cite{jeff22} for solid ellipsoidal objects,
where $\phi$ is kept in the original position.
These results show good agreement with recent experiments of lipid vesicles \cite{kant06,desc09,desc09a}.

\begin{figure}
\includegraphics{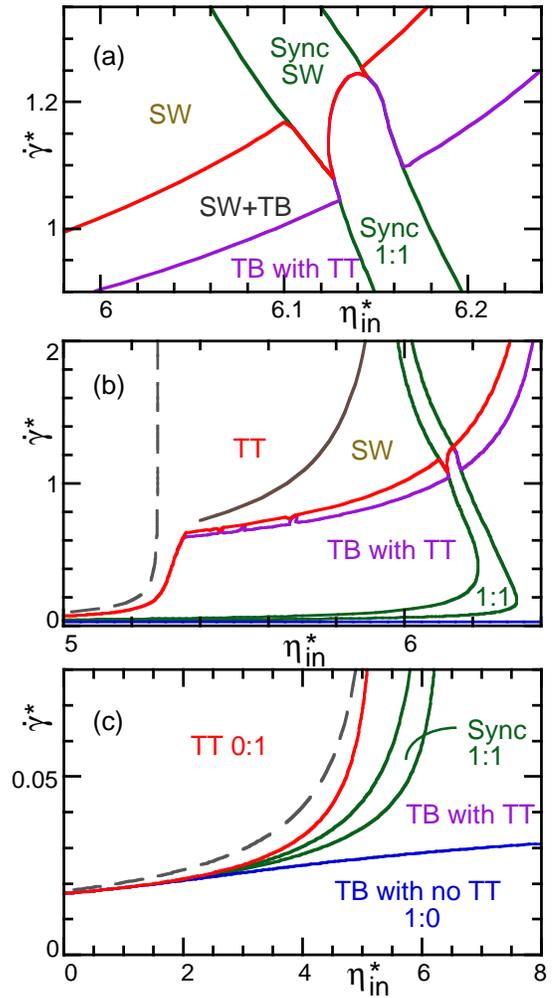}
\caption{ \label{fig:phase_e1}
(Color online)
Phase diagram ($\eta_{\rm {in}}^*,\dot\gamma^*$) of microcapsules at $k_{\rm e}=1$.
(a) Enlarged figure for the region of synchronized rotation.
(b) Region at high $\eta_{\rm {in}}^*$.
(c) Region at low $\dot\gamma^*$. 
Solid lines represent the phase boundaries.
In the right region of dashed line, the vesicle passes through or remains at $\theta<0$.
Numbers represent the rotation frequencies $f_{\rm {rot}}^{\theta}:f_{\rm {rot}}^{\phi}$.
}
\end{figure}

\begin{figure}
\includegraphics{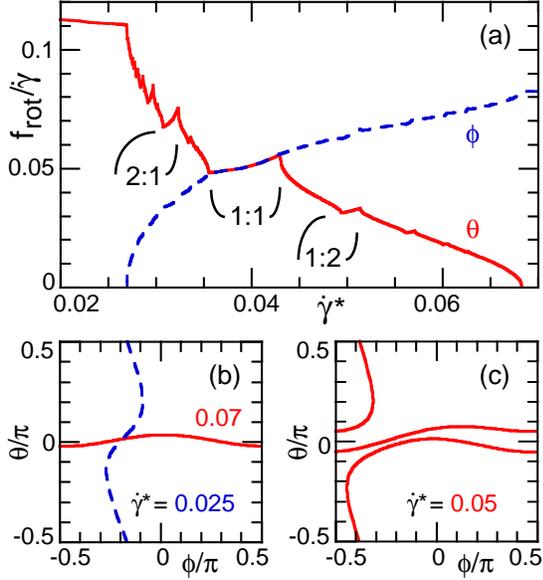}
\caption{ \label{fig:e1z5}
(Color online)
Dynamic mode transition from TB to TT phase at $k_{\rm e}=1$ and $\eta_{\rm {in}}^*=5$.
(a) Rotation frequencies $f_{\rm {rot}}$ of 
$\theta$ (solid line) and $\phi$ (dashed line).
(b) Trajectories of TB with $f_{\rm {rot}}^{\theta}:f_{\rm {rot}}^{\phi}=1:0$ 
and TT with $f_{\rm {rot}}^{\theta}:f_{\rm {rot}}^{\phi}=0:1$ 
at $\dot\gamma^*=0.025$ (dashed line) and $0.07$ (solid line), respectively.
(c) Trajectory of TB accompanied by TT $\phi$ rotation 
with $f_{\rm {rot}}^{\theta}:f_{\rm {rot}}^{\phi}=1:2$ 
at $\dot\gamma^*=0.05$.
}
\end{figure}

\begin{figure}
\includegraphics{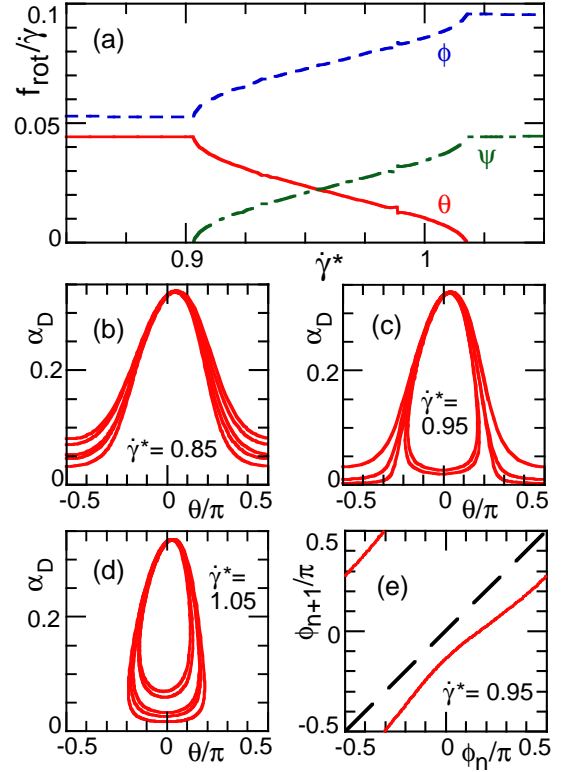}
\caption{ \label{fig:e1z6}
(Color online)
Dynamic mode transition from TB to SW phase at $k_{\rm e}=1$ and $\eta_{\rm {in}}^*=6$.
(a) Rotation frequencies $f_{\rm {rot}}$ of $\theta$ (solid line), $\phi$ (dashed line), 
and swinging $\psi$  (dashed-dotted line).
(b) Trajectory at $\dot\gamma^*=0.85$: TB accompanied by TT $\phi$ rotation.
(c) Trajectory at $\dot\gamma^*=0.95$: Coexistence of TB and SW motions.
(d) Trajectory at $\dot\gamma^*=1.05$: SW motion.
(e) Return map at $\dot\gamma^*=0.95$ for
 the angle $\phi$ when trajectory crosses $\theta= 0$ at $\alpha_{\rm D}>0.2$.
Dashed line represents $\phi_{n+1}=\phi_n$.
}
\end{figure}

\begin{figure}
\includegraphics{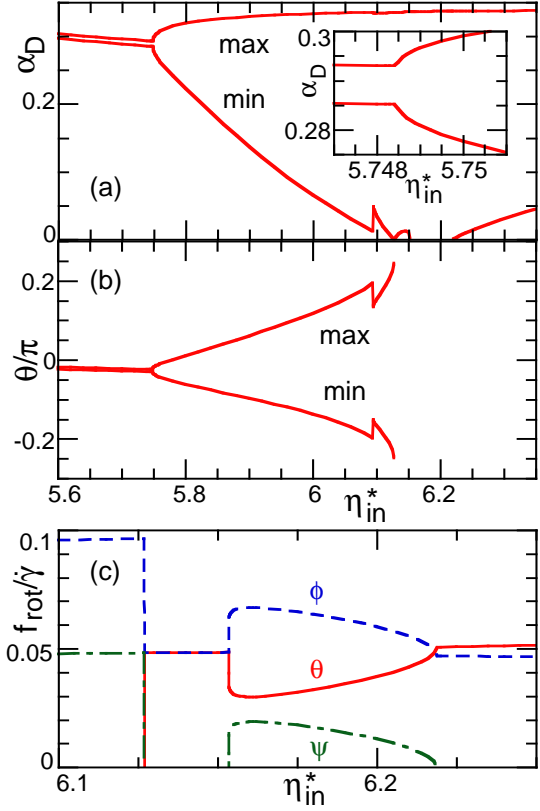}
\caption{ \label{fig:e1g12}
(Color online)
Dynamic mode transitions from TT, SW to TB phase via synchronized region 
at $k_{\rm e}=1$ and $\dot\gamma^*=1.2$: TT for $\eta_{\rm {in}}^*<5.7484$,
SW for $5.7484<\eta_{\rm {in}}^*<6.127$, SW+TB for $6.154<\eta_{\rm {in}}^*<6.219$,
and TB with TT for $6.127<\eta_{\rm {in}}^*<6.154$ and $\eta_{\rm {in}}^*>6.219$.
Maximum and minimum of (a) $\alpha_{\rm D}$ and (b) $\theta$.
The region around transition between TT and SW is enlarged in the inset in (a).
(c) Rotation frequencies $f_{\rm {rot}}$ of $\theta$ (solid line), $\phi$ (dashed line), 
and swinging $\psi$  (dashed-dotted line).
}
\end{figure}

\begin{figure}
\includegraphics{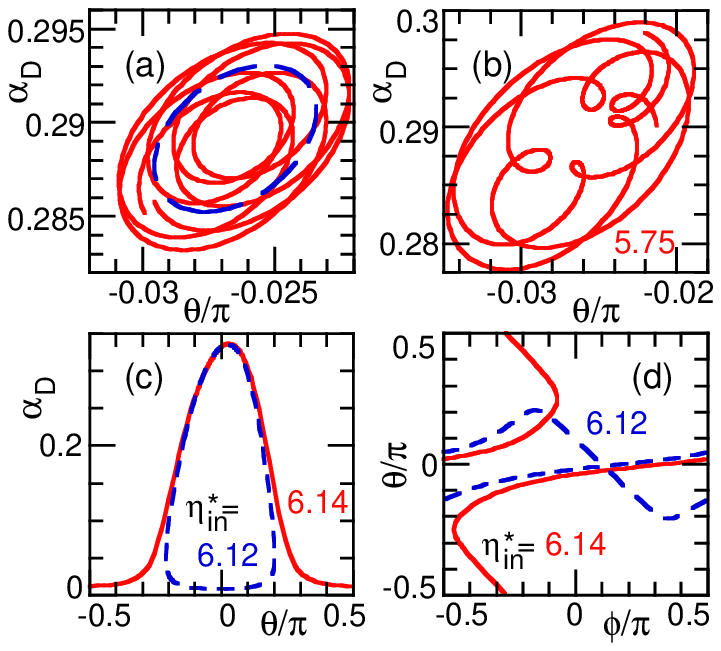}
\caption{ \label{fig:e1g12rq}
(Color online)
Trajectories of microcapsules at $k_{\rm e}=1$ and $\dot\gamma^*=1.2$.
(a) 
Dashed and solid lines represent trajectories in the TT phases (type II oscillation) at $\eta_{\rm {in}}^*=5.748$
and in the SW phase (unsynchronized type I and II oscillations) at $\eta_{\rm {in}}^*=5.7486$, respectively.
(b) Unsynchronized oscillations with larger amplitude of the type I oscillation
 in the SW phase at $\eta_{\rm {in}}^*=5.75$.
(c--d) 
Synchronized rotations in the SW phase at $\eta_{\rm {in}}^*=6.12$ (dashed line) and 
in the TB phase at $\eta_{\rm {in}}^*=6.14$ (solid line).
}
\end{figure}

\subsection{Microcapsule with low shear elasticity} \label{sec:rel1}

When the membrane shear elasticity is added, the dynamics is changed
from the above dynamics of fluid vesicles.
First, we consider the dynamics of microcapsules at $k_{\rm e}=1$,
where shear elasticity is relatively lower than the bending elasticity.
The energy barrier for $\phi$ rotation is $F^*=0.21$ at $\phi=\pi/2$,
which is five times lower than the energy barrier $F^*=0.97$ for the capsule shrinkage at $\alpha_{\rm D}=0$.
Because of this energy scale difference,
the phase diagram can be separated into two regions:
high shear region $\dot\gamma^* \gtrsim 0.8$ shown in Figs. \ref{fig:phase_e1}(a) and (b),
 and low shear region $\dot\gamma^* \lesssim 0.1$ shown in Fig. \ref{fig:phase_e1}(c).

\subsubsection{TB-TT transition} \label{sec:rel1tt}

At low shear rate $\dot\gamma^* \lesssim 0.1$,
the shape deformation is very small so microcapsules
keep their prolate shape, i.e.  $\alpha_{\rm D}\simeq \alpha_0$.
Thus, the generalized KS theory for fix-shaped capsules proposed by Skotheim and Secomb \cite{skot07}
(Eq. (\ref{eq:qks}) and Eq. (\ref{eq:phiks2}) with constant $\alpha_{\rm D}$)
predicts the phase behavior in this region very well.
At very low shear rate  $\dot\gamma^* \lesssim 0.02$,
the capsules show TB motion with no TT unlike fluid vesicles; see dashed line in Fig. \ref{fig:e1z5}(b).
The energy barrier prevents $\phi$ rotation.
Without TT rotation, the whole capsule has to rotate (tumble) 
due to the shear torque applied on the capsule surface.

As $\dot\gamma^*$ increases, $\phi$ starts to rotate
and then the TT frequency $f_{\rm {rot}}^{\phi}$ gradually increases
and the TB frequency $f_{\rm {rot}}^{\theta}$ decreases; see Fig. \ref{fig:e1z5}(a).
Here, an angle change of $\pi$ is counted as one rotation.
In this coexistence range of two rotations (TB with TT),
the synchronization of $\phi$ and $\theta$ can occur
with the integer ratios of $f_{\rm {rot}}^{\theta}$ and $f_{\rm {rot}}^{\phi}$.
This infinite number of synchronization phases is called Devil's staircase \cite{berg84}.
The trajectory of the synchronized rotations with $f_{\rm {rot}}^{\theta}:f_{\rm {rot}}^{\phi}=1:2$ are shown in Fig. \ref{fig:e1z5}(c). 

At sufficiently large $\dot\gamma^* > \dot\gamma^*_{\rm {tt}}$ 
($\dot\gamma^*_{\rm {tt}}=0.068$ at $\eta_{\rm {in}}^*=5$),
$\theta$ stops rotation and the TT phase appears.
With the frequency of $\phi$ rotation, $\theta$ and $\alpha_{\rm D}$ oscillate (swing)
around the steady values for fluid vesicles ($k_{\rm e}=0$).
As $\dot\gamma^*$ increases, the oscillation amplitudes decrease \cite{abka07,nogu09b}.
In this paper, we call this phase TT following the phase diagram of fluid vesicles
(it is also called SW \cite{abka07,kess08,sui08} or OS \cite{bagc09} in the previous papers).
As $\dot\gamma^* \to \dot\gamma^*_{\rm {tt}}$,
the mean angle $\langle\theta\rangle$ approaches to null
and the $\theta$ amplitude increases and then 
$\theta$ passes through $\theta<0$ during this oscillation;
see solid line for $\dot\gamma^*=0.07$ in  Fig. \ref{fig:e1z5}(b).
This crossing region appears 
between the solid (red) and dashed lines in  Fig. \ref{fig:phase_e1}(c).
This dashed line does not represent a phase boundary.
In crossing this dashed line,
 the averages and amplitudes of $\theta$ and $\alpha_{\rm D}$  show 
smooth change \cite{nogu09b}. No discrete changes of trajectories are seen.
This trajectory crossing $\theta<0$ resembles the one in the SW phase of fluid vesicles.
However, they are caused by a different mechanism. 
We call them type I and type II  (swinging) oscillations.
Type I: Oscillation induced by the shape deformation.
For the oscillation of fluid vesicles, the large shape deformation is essential as explained in Sec. \ref{sec:rfluid}.
Type II: Oscillation induced by the energy barrier to $\phi$ rotation.
This oscillation can be reproduced with a fix-shaped approximation
including the passage through the negative angle $\theta$. The resulting oscillation of the $\phi$ rotational velocity
causes $\theta$ and $\alpha_{\rm D}$ oscillations.
These two types of oscillations cannot be distinguished only by trajectory shapes.

\subsubsection{SW phase} \label{sec:rel1sw}

The microcapsule at $k_{\rm e}=1$ has a SW phase around the region of the SW phase at $k_{\rm e}=0$;
compare Figs. \ref{fig:phase_e0}(a) and  \ref{fig:phase_e1}(b).
In the SW phase, the type I oscillation occurs in addition to the type II oscillation.
It is found that this coexistence makes
the transition from TB with TT to SW a gradual change; see Fig. \ref{fig:e1z6}.
The SW $\psi$ frequency is calculated as a frequency of crossing
$\theta=\theta_{\rm {mid}}$ upwards at $\alpha_{\rm D}<\alpha_{\rm {mid}}$ in Fig. \ref{fig:e1z6},
where $(\theta_{\rm {mid}}, \alpha_{\rm {mid}})$ is a point inside the SW trajectory,
and $(\theta_{\rm {mid}}, \alpha_{\rm {mid}})= (0,0.2)$ is chosen here.
The type II oscillation generates the fluctuations of the TB or SW trajectories,
when the oscillations and $\phi$ rotation are not synchronized; 
see return map in Fig. \ref{fig:e1z6}(e).
As $\dot\gamma^*$ increases, an unsynchronized (intermittent) SW oscillation starts,
and then its frequency increases 
while the total of the TB and SW frequencies is almost constant.

The TT-SW transition is slightly modified from fluid vesicles.
Both TT and SW phases are accompanied by the type II oscillation.
The transition point is detected as a discrete point of the slope
of the amplitudes of $\alpha_{\rm D}$ or $\theta$ with respect to $\eta_{\rm {in}}^*$;
see the region around the transition point $\eta_{\rm {in}}^*=5.7484$ in Figs. \ref{fig:e1g12}(a) and (b).
Two types of oscillations without synchronization are shown in
 Figs. \ref{fig:e1g12rq}(a) and (b).

Synchronization of type I and type II oscillations can occur 
with integer ratio of the oscillation frequencies.
The synchronization with $f_{\rm {rot}}^{\theta}:f_{\rm {rot}}^{\phi}:f_{\rm {rot}}^{\psi}=0:2:1$
has the widest region (denoted by `Sync SW' in  Fig. \ref{fig:phase_e1}(a)).
Two $\phi$ rotations occur every $\theta$ oscillation in this region; see dashed lines in Figs. \ref{fig:e1g12rq}(c) and (d).
The TB-SW transition in synchronized region  ($\eta_{\rm {in}}^*=6.127$ at $\dot\gamma^*=1.2$)
is a discrete transition like that of the fluid vesicles;
The limit cycle with  $f_{\rm {rot}}^{\theta}:f_{\rm {rot}}^{\phi}:f_{\rm {rot}}^{\psi}=0:2:1$ becomes
 $f_{\rm {rot}}^{\theta}:f_{\rm {rot}}^{\phi}:f_{\rm {rot}}^{\psi}=1:1:0$.
At the boundaries of the synchronized region ($\eta_{\rm {in}}^*=6.094$ and $6.154$ at $\dot\gamma^*=1.2$),
the trajectories show discrete changes; see Figs. \ref{fig:e1g12}(a--c).
Other synchronized regions with different ratios of $f_{\rm {rot}}^{\theta}:f_{\rm {rot}}^{\phi}$
are seen as kinks or slits of the boundary line between 'SW+TB' and 'TB with TT' phases in Fig. \ref{fig:phase_e1}(b).

\begin{figure}
\includegraphics{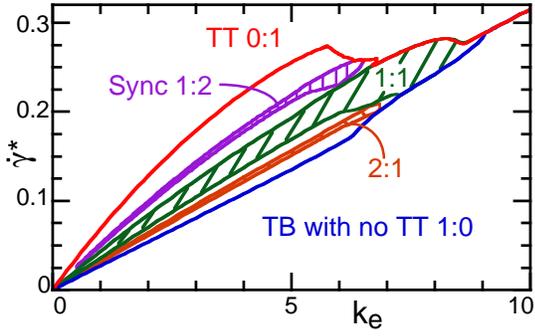}
\caption{ \label{fig:phase_z5}
(Color online)
Phase diagram ($k_{\rm e},\dot\gamma^*$) of microcapsules
at $\eta_{\rm {in}}^*=5$.
Top and bottom lines represent the phase boundaries of TT ($f_{\rm {rot}}^{\theta}:f_{\rm {rot}}^{\phi}=0:1$) and 
TB with no TT ($f_{\rm {rot}}^{\theta}:f_{\rm {rot}}^{\phi}=1:0$), respectively.
Shadowed regions represent synchronized phase with $f_{\rm {rot}}^{\theta}:f_{\rm {rot}}^{\phi}=1:2$, $1:1$, and $2:1$.
}
\end{figure}

\begin{figure}
\includegraphics{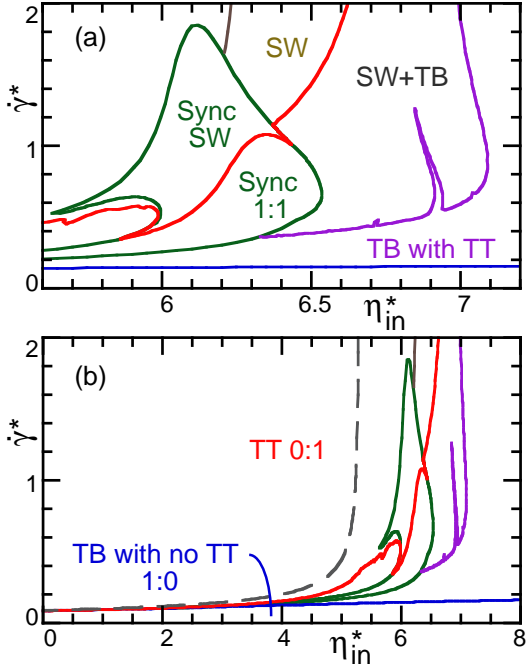}
\caption{ \label{fig:phase_e5}
(Color online)
Phase diagram ($\eta_{\rm {in}}^*,\dot\gamma^*$) of microcapsules at $k_{\rm e}=5$.
Lines represent the phase boundaries and the region with $\theta<0$
as depicted in Fig. \ref{fig:phase_e1}.
}
\end{figure}

\begin{figure}
\includegraphics{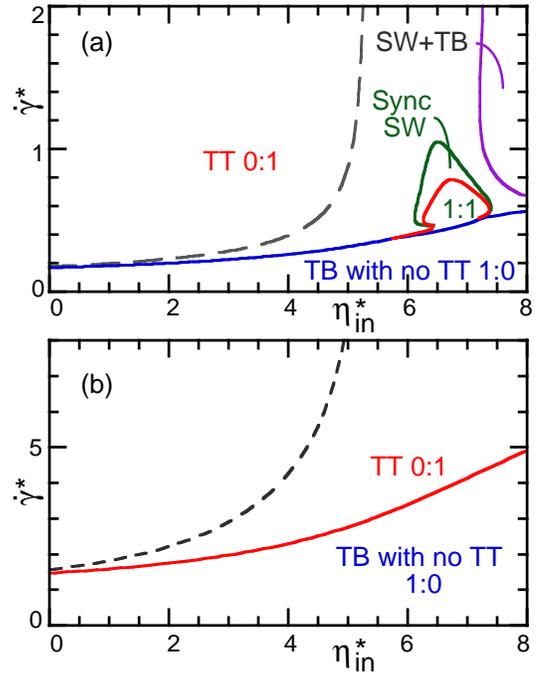}
\caption{ \label{fig:phase_e10}
(Color online)
Phase diagram ($\eta_{\rm {in}}^*,\dot\gamma^*$) of microcapsules at (a) $k_{\rm e}=10$ and (b) $k_{\rm e}=100$.
Lines represent the phase boundaries and the region with $\theta<0$
as depicted in Fig. \ref{fig:phase_e1}.
}
\end{figure}

\subsection{Microcapsule with high shear elasticity} \label{sec:rel5}

With increasing $k_{\rm e}$, the phase diagram is found to show remarkable changes.
Phase diagrams at $\eta_{\rm {in}}^*=5$ and at $k_{\rm e}=5$, $10$, and $100$
are shown in Figs. \ref{fig:phase_z5}--\ref{fig:phase_e10}.
At $k_{\rm e}\lesssim 5$, the TT-TB transition shear rates $\dot\gamma^*$
are approximately linear to $k_{\rm e}$,
as the height of the energy barrier for $\phi$ rotation linearly increases;
see Fig. \ref{fig:phase_z5}.
A free-energy minimum at $\phi=\pi/2$ (or $\alpha_{\rm D}<0$ at $\phi=0$) 
disappears at $k_{\rm e} > k_{\rm e}^{\rm c}=8.2$; see Fig. \ref{fig:pot}.
Around this threshold $k_{\rm e} \simeq k_{\rm e}^{\rm c}$, the phase behavior dramatically changes.
At $k_{\rm e} =5$ (slightly below $k_{\rm e}^{\rm c}$),
the TB-TT transition line becomes close to the SW phase boundary as shown in Fig. \ref{fig:phase_e5}.
The larger amplitudes of type II oscillation make 
the coexistence region 'SW +TB' wider.
The synchronized region with $f_{\rm {rot}}^{\theta}:f_{\rm {rot}}^{\phi}=1:1$ at the TB-SW boundary
 also grows from $\Delta \eta_{\rm {in}}^*=0.04$ ($k_{\rm e}=1$) 
to $\Delta \eta_{\rm {in}}^*=0.4$ ($k_{\rm e}=5$).
A slit at $\eta_{\rm {in}}^*=6.9$ in Fig. \ref{fig:phase_e5}
shows a synchronized region with $f_{\rm {rot}}^{\theta}:f_{\rm {rot}}^{\phi}=3:2$.
Narrower slits with different ratios of $f_{\rm {rot}}^{\theta}:f_{\rm {rot}}^{\phi}$ are omitted
in the phase diagram in Fig. \ref{fig:phase_e5}.

With a further increase in $k_{\rm e}$,
the region width of 'TB with TT' decreases, and then vanishes; see Figs. \ref{fig:phase_z5} and \ref{fig:phase_e10}.
At $\eta_{\rm {in}}^*=5$, first the regions of unsynchronized (intermittent) rotation are reduced, and
the synchronized region with $f_{\rm {rot}}^{\theta}:f_{\rm {rot}}^{\phi}=1:1$ becomes  
dominant. At $k_{\rm e} \simeq 8$, this synchronized region only remains for 'TB with TT'.
At $k_{\rm e} > 9$, direct transition from 'TB with no TT'  to TT phase occurs.

\begin{figure}
\includegraphics{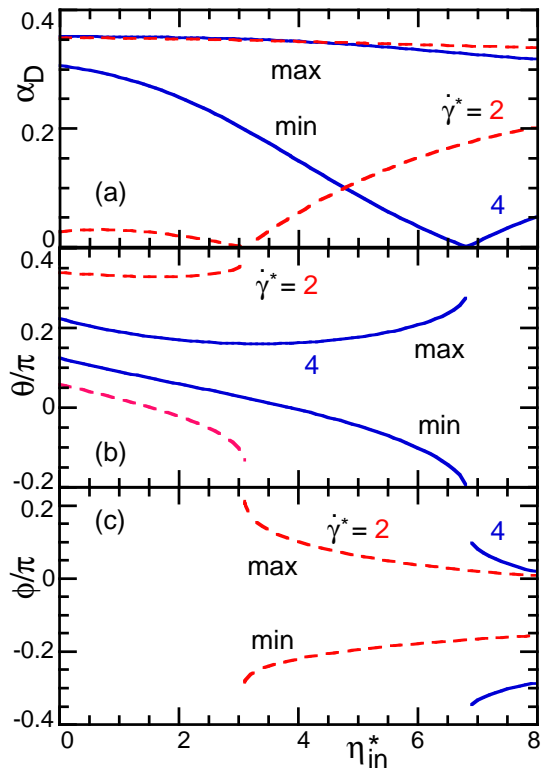}
\caption{ \label{fig:rq_e100}
(Color online)
Viscosity ratio $\eta_{\rm {in}}^*$ dependence of
maximum and minimum of (a) $\alpha_{\rm D}$, (b) $\theta$, and (c) $\phi$ at $k_{\rm e}=100$.
Dashed and solid lines represent $\dot\gamma^*=2$ and $4$, respectively.
}
\end{figure}

At $\eta_{\rm {in}}^*=100$, only two phases, TT with $f_{\rm {rot}}^{\theta}:f_{\rm {rot}}^{\phi}=0:1$ and TB with  $f_{\rm {rot}}^{\theta}:f_{\rm {rot}}^{\phi}=1:0$, remain; see Fig. \ref{fig:phase_e10}(b).
This phase diagram agrees well with recent simulations for microcapsules with high shear elasticity 
at $\gamma=100$~\cite{kess08} and $\gamma\to \infty$ \cite{bagc09}. 
The rotations of $\phi$ and $\theta$ do not occur simultaneously at any point in the phase diagram.
During tumbling, the shrinking force caused by shear elasticity
prevents a passage through $\phi=\pi/2$.
In the TT phase, the shear force $\propto \dot\gamma \sin(2 \theta)$ allows the $\phi$ rotation through  $\phi=\pi/2$.
The transition between these two phases occurs when the trajectory reaches $\alpha_{\rm D}=0$
like the SW-TB transition of fluid vesicles; see Fig. \ref{fig:rq_e100}.
In the TT phase,  the amplitudes of $\theta$ and $\alpha_{\rm D}$ oscillations smoothly increase
with increasing $\eta_{\rm {in}}^*$; 
$\theta$ and $\alpha_{\rm D}$ oscillate with the TT frequency 
($f_{\rm {rot}}^{\theta}:f_{\rm {rot}}^{\phi}:f_{\rm {rot}}^{\psi}=0:1:1$) as in the TT phase at smaller $k_{\rm e}$.
At small and large $\eta_{\rm {in}}^*$, type II and type I oscillations may be dominant, respectively.
However, a clear boundary of TT and SW phases is not found.
The dashed line in Fig. \ref{fig:phase_e10}(b), 
where the minimum of $\theta$ crosses $\theta=0$, is not a phase boundary either.

\begin{figure}
\includegraphics{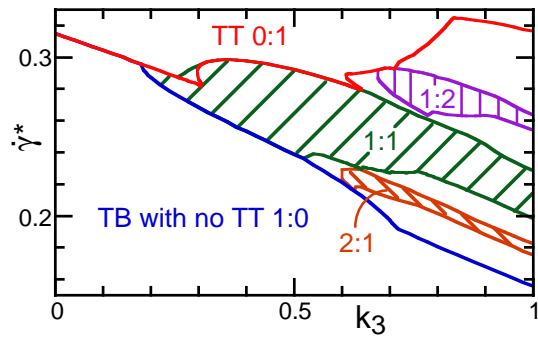}
\caption{ \label{fig:phase_k3}
(Color online)
Phase diagram ($k_3,\dot\gamma^*$) of microcapsules
at $\eta_{\rm {in}}^*=5$ and $k_{\rm e}=10$.
Lines and shadowed regions represent the phase boundaries and 
synchronized phase, respectively, as depicted in Fig. \ref{fig:phase_z5}.
The coexistence region of two limit cycles TT ($f_{\rm {rot}}^{\theta}:f_{\rm {rot}}^{\phi}=0:1$) 
and TB with TT ($f_{\rm {rot}}^{\theta}:f_{\rm {rot}}^{\phi}=1:1$)
is seen around $(k_3, \dot\gamma^*)=(0.3, 0.3)$.
}
\end{figure}

\subsection{Effects of a saddle point of the potential} \label{sec:min}

As shown above,
the phase diagram of microcapsules is changed at $k_{\rm e}\simeq k_{\rm c}$,
where the saddle point in the free energy
(the energy minimum at $\alpha_{\rm D}<0$ for $\phi=0$ 
and the energy maximum in the $\phi$ rotation for constant $\alpha_{\rm D}$) vanishes.
To check whether the saddle point is essential for the phase behavior
or it is an accidental coincidence,
the higher-order term $F_3$ is added to the potential.
This term restores the energy minimum at $\alpha_{\rm D}<0$,
while the energy profile is little modified at $\alpha_{\rm D}>0$ as shown in Fig. \ref{fig:pot}.
Figure \ref{fig:phase_k3} shows that the synchronized and intermittent regions in the phase diagram
reappear with an increase in $k_3$.
The resulting dynamics are very similar to those at low $k_{\rm e}$ with $k_3=0$.
Thus, the free-energy landscape for $\phi$ rotation is a key feature to determine the dynamics of microcapsules.

\section{Discussion}~\label{sec:dis}

Our results revealed that a saddle point of the potential
changes the phase diagram of microcapsules.
For prolate capsules with linear-elastic membrane at $V^*=0.9$,
the threshold of the phase diagram $k_{\rm e}\sim 10$ corresponds to
F\"oppl-von K\'arm\'an number $\gamma=Y_{\rm {mb}} R_0^2/\kappa = 1 \sim 10$,
which is lower than those of typical synthetic microcapsules ($\gamma \gtrsim 100$) 
and RBCs ($\gamma\simeq 400$).
However, the existence of a saddle point also depends on capsule shapes and buckling instability.
A saddle point is observed for RBCs by experiments \cite{fisc04} and simulations \cite{nogu09b}
although RBCs have large $\gamma$.
When an RBC is compressed
in the $x$ direction ($y$ axis is the symmetric axis of a discocyte) with a constraint of $\phi=0$,
two dimples are buckled and new dimples appear on the $x$ axis.
These dimples generate an energy minimum at $\phi=0$ (saddle point).
Therefore, RBCs exhibit synchronized or intermittent rotations of $\phi$ and $\theta$ \cite{nogu09b}.

\begin{figure}
\includegraphics{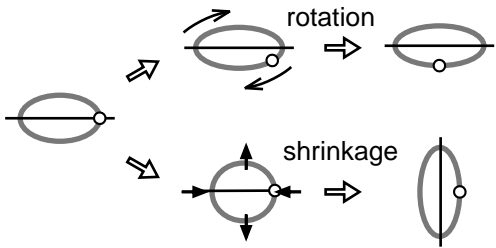}
\caption{ \label{fig:cart2}
Schematic representation of two relaxation pathways from $(\alpha_{\rm D},\phi)=(\alpha_0,\pi/2)$
to $(\alpha_{\rm D},\phi)=(\alpha_0,0)$ after stopping shear flow.
}
\end{figure}

Fischer reported two types of relaxation dynamics for RBCs after the shear stops \cite{fisc04}.
After TT motions in steady shear flow, the shear is stopped at $\phi \simeq \pi/2$.
Then the RBCs return to their thermal equilibrium state by $\phi$ rotation or shrinkage;
see Fig. \ref{fig:cart2}.
During the shrinkage, the buckling is observed.
For microcapsules, the existence of a saddle point of the potential can be checked by 
the observation of this relaxation.
If capsules have no saddle point, it would relax by continuous shrinkage (not by $\phi$ rotation).
Buckling or $\phi$ rotation suggests an energy barrier for the shrinkage.
For synthetic capsules, a saddle point can be induced by buckling and nonlinear shear elasticity.
At large $\gamma \gtrsim 150$, buckling transition can occur for spherical and ellipsoidal capsules  \cite{lidm03}.
Hysteresis at the buckling transition was reported in the simulation of microcapsules with $\gamma\to \infty$ for elongational flow \cite{dods08}.
Similar buckling may appear in shear flow.

We assumed that the major axis of the microcapsule is on the vorticity ($xy$) plane.
This assumption is valid in most of the parameter range.
Lebedev {\it et al.} reported spinning motion, where the major axis rotates out of the  vorticity plane,
at very large $\dot\gamma^*$ and large $\eta_{\rm {in}}^*$ 
($\dot\gamma^* \gtrsim 2$ and $\eta_{\rm {in}}^*\gtrsim 8$ at $V^*=0.9$)\cite{lebe08}.
Thus, spinning dynamics of microcapsules is also expected at large $\eta_{\rm {in}}^*$ 
but it is beyond the scope of our present study.

Here, we considered membrane with constant surface area.
This condition is valid for lipid vesicles and cells,
because the lipid membrane is almost incompressible.
Compared to them, polymer membranes are area compressible.
For microcapsules with a prolate shape,
the effects of area expansion can be  included by a slight modification
of $F$ and factors $f_i$ for $0 \leq i \leq 3$ in Eqs. (\ref{eq:ald}), (\ref{eq:qks}), and (\ref{eq:phiks2}).
For a spherical microcapsule in the absence of flow, however,
area expansion is essential,
since no deformation from a sphere is allowed without area or volume change.
The spherical capsules have energy minimum at $\alpha_{\rm D}=0$
so that no saddle point appear even at $\mu=0$.
Thus, the spherical capsules likely do not have `TB with TT' phase.

Capsules with $V^*\lesssim 0.7$ may have multiple minima in the free-energy landscape;
they would show more complicated phase behavior.
Fluid vesicles with $V^*\simeq 0.6$ have 
three (meta-)stable states: discocyte, prolate, and stomatocyte \cite{seif97}.
Shear flow induces the elongating shape transition from  discocyte or stomatocyte to prolate
as well as the shrinking transition from prolate to discocyte \cite{nogu04,nogu05}.
The transition dynamics between discocyte and prolate can be well reproduced by two differential equations
of shape parameter and inclination angle $\theta$ similar to Eqs. (\ref{eq:ald}) and (\ref{eq:qks2}).
However, the transition dynamics from stomatocyte to prolate cannot be reproduced 
by the above two equations. At least two shape parameters are necessary to describe
the direct shape change (not via discocyte) obtained in our previous simulations \cite{nogu05}.
When shear elasticity is added to discocyte vesicles,
these metastable states become less stable.
In our previous paper, we investigated the dynamics of RBCs at low shear rate 
by dynamic equations with one shape parameter \cite{nogu09b}.
At large shear rate with large $\eta_{\rm {in}}^*$, RBCs can become $\alpha_{\rm D} \simeq 0$,
where one shape parameter may not be sufficient to describe the RBC shape.

\section{Summary}~\label{sec:sum}

We have systematically studied the dynamics of prolate microcapsules in steady shear flow.
The capsules exhibit four types of motions,
tank-treading, tumbling, and two types of swinging oscillations.
When the potential has a saddle point at the phase angle $\phi=\pi/2$ (low shear elasticity), 
these motions can occur simultaneously with or without synchronization.
As shear rate increases at small viscosity ratio $\eta_{\rm {in}}^*$, 
the capsules show the transition from tumbling with no tank-treading membrane rotation
to tank-treading mode via coexistence of tumbling and tank-treading rotations.
Swinging phase appears at large $\eta_{\rm {in}}^*$  and large shear rate $\dot\gamma^*$,
where two type of swinging oscillations coexist.
When the potential has no saddle point (high shear elasticity),
the phase diagram is drastically changed.
The coexistence phases disappear and only two phases (tank-treading and tumbling phases with no tank-treading)
remain.

We have investigated only prolate capsules with a homogeneous membrane in this paper.
However, our model can be applied easily to other capsules by the modification of capsule potential $F$.
Thus, it can cover a wide range of soft objects from fluid vesicles to microcapsules and cells.

The coexistence of tumbling and tank-treading rotations has not yet been observed in experiments and simulations.
Our study predicts it would occur for low shear-elastic capsules or capsules with dimples like RBCs.
Further experiments and simulations are expected to confirm our predictions.

\begin{acknowledgments}
This study is supported by a Grant-in-Aid for Scientific Research on Priority Area ``Soft Matter Physics'' from
the Ministry of Education, Culture, Sports, Science, and Technology of Japan.
\end{acknowledgments}

\begin{appendix}
\section{Keller-Skalak Theory}

In the KS theory, a vesicle is assumed to have a fixed ellipsoidal shape, 
\begin{eqnarray}
\Big(\frac{x_1}{a_1}\Big)^2 +\Big(\frac{x_2}{a_2}\Big)^2 +\Big(\frac{x_3}{a_3}\Big)^2 =1, 
\end{eqnarray}
where $a_i$ denote the semi-axes of the ellipsoid, and
the coordinate axes $x_i$ point along its principal directions. 
The $x_1$ and $x_2$ axes, with $a_1>a_2$, are on the vorticity ($xy$) plane,
and the $x_3$ axis is  in the vorticity ($z$) direction.
The maximum lengths in three directions are $L_1=2a_1$, $L_2=2a_2$, and $L_3=2a_3$.
The velocity field on the membrane is assumed to be 
\begin{eqnarray}
{\bf v}^{\rm {m}}=
               \omega \Big(-\frac{a_1}{a_2}x_2,\frac{a_2}{a_1}x_1,0\Big).
\label{eq:KS-vel}
\end{eqnarray}
Equations (\ref{eq:qks}) and (\ref{eq:phiks}) of vesicle motion are derived from the energy balance 
between supply from the external fluid of the vesicle and dissipation inside the vesicle and on the membrane.
The factors $f_0$, $f_1$, $f_2$, and $f_3$ are given by
\begin{eqnarray*}
f_0 &=& 2/(a_1/a_2+a_2/a_1) = (1-\alpha_{\rm D}^2)/(1+\alpha_{\rm D}^2),\\
f_1 &=& 0.5(a_1/a_2-a_2/a_1) = 2\alpha_{\rm D}/(1-\alpha_{\rm D}^2),\\
f_2 &=& 0.5g(\alpha_1^2+\alpha_2^2),\\
f_3 &=& 0.5E_{\rm s}R_0/(f_1^2V),\\
g &=& \int_0^\infty (\alpha_1^2+s)^{-3/2}(\alpha_2^2+s)^{-3/2}
          (\alpha_3^2+s)^{-1/2}ds,\\
\alpha_i &=& a_i/(a_1a_2a_3)^{1/3},
\end{eqnarray*}
where $E_{\rm s}$ is an integral of shear stress over the membrane surface \cite{tran84,nogu05}.
At $V^*=0.9$, we numerically obtained 
$f_2= 0.3185+0.6 \alpha_{\rm D}^2 +14 \alpha_{\rm D}^6 + \exp\{124 (\alpha_{\rm D}-0.39)\}$.

\section{Estimation of shear elastic energy}

For quasi-spherical capsules,
the capsule shape 
can be expanded in spherical harmonics $Y_{l,m}$ as 
$R= R_{\rm V}(1+\sum_{l,m}u_{l,m}Y_{l,m})$. 
Ellipsoidal shapes are described with  $l=0, 2$ modes as
$R= R_{\rm V}\{1-\Delta_{\rm S}/8\pi + u_{2,2}(Y_{2,2}+Y_{2,-2}) +u_{2,0}Y_{2,0}\}$ $=$
$R_{\rm V}\{1-\Delta_{\rm S}/8\pi + \sqrt{15/8\pi}\sin^2(\theta)\cos(2\varphi)u_{2,2} + (\sqrt{5/\pi}/4)(3\cos^2(\theta)-1)u_{2,0} \}$ 
with $u_{2,0}^2+2u_{2,2}^2=\Delta_{\rm S}/2$ in the spherical coordinate representation $(R,\theta,\varphi)$.
A prolate shape with $u_{2,0}=u_{2,0}^{\rm {eq}}=-\sqrt{\Delta_{\rm S}/2}/2$ and $u_{2,2}=u_{2,2}^{\rm {eq}}=\sqrt{3\Delta_{\rm S}}/4$ is chosen as a rest state in the absence of flow.
The displacement is assumed in the radial direction: ${\bf u}=u_r {\bf e}_r$ with $u_r= (u_{2,2}-u_{2,2}^{\rm {eq}})(Y_{2,2}+Y_{2,-2}) +(u_{2,0}-u_{2,0}^{\rm {eq}})Y_{2,0}$.
The shear elastic energy $\mu\int (\varepsilon_{1,1}-\varepsilon_{2,2})^2/2+2{\varepsilon_{1,2}}^2  dS$ is numerically calculated,
where $\varepsilon_{i,j}$ is the two-dimensional strain tensor on the membrane surface.

\end{appendix}


\end{document}